\documentclass[twocolumn, trackchanges]{aastex631}

\usepackage{subfigure}
\usepackage{amsmath}
\usepackage{color}
\usepackage{soul}
\usepackage{blindtext}
\usepackage{fancybox}
\usepackage{listings}

\graphicspath{{./}{figures/}}

\shorttitle{Hydrogenated carbonaceous molecular particles}
\shortauthors{Ricca~et~al.}

\received{XXX, 2024}
\revised{XXX, 2024}

\submitjournal{ApJ}

\begin{document}

\title{The \textit{JWST} Proto-PAH project. Computational modeling of the emission carriers}

\author[0000-0002-3141-0630]{A.~Ricca}
\affiliation{NASA Ames Research Center, MS 245-6, Moffett Field, CA 94035-1000, USA}
\affiliation{Carl Sagan Center, SETI Institute, 339 Bernardo Avenue, Suite 200, Mountain View, CA 94043, USA}

\author[0000-0003-4520-1044]{G.~C.\ Sloan}
\affiliation{Space Telescope Science Institute, 3700 San Martin Drive, 
  Baltimore, MD 21218, USA}
\affiliation{Department of Physics and Astronomy, University of North 
  Carolina, Chapel Hill, NC 27599-3255, USA}

\author[0000-0002-2541-1602]{E.~Peeters}
\affiliation{Department of Physics and Astronomy, The University of Western Ontario, London, ON N6A 3K7, Canada}
\affiliation{The Institute for Earth and Space Exploration, The University of Western Ontario, London, ON N6A 3K7, Canada}

\author[0000-0002-2541-1602]{J.~Cami}
\affiliation{Department of Physics and Astronomy, The University of Western Ontario, London, ON N6A 3K7, Canada}
\affiliation{The Institute for Earth and Space Exploration, The University of Western Ontario, London, ON N6A 3K7, Canada}

\author[0009-0009-4643-2734]{N.~Clark}
\affiliation{Department of Physics and Astronomy, The University of Western Ontario, London, ON N6A 3K7, Canada}
\affiliation{The Institute for Earth and Space Exploration, The University of Western Ontario, London, ON N6A 3K7, Canada}

\author[0000-0002-5529-5593]{M.~Matsuura}
\affiliation{Cardiff Hub for Astrophysics Research and Technology (CHART), PACES, Cardiff University, The Parade, Cardiff CF24 3AA, UK}

\author[0000-0002-6858-5063]{R.~Sahai}
\affiliation{Jet Propulsion Laboratory, California Institute of Technology, 4800 Oak Grove Dr., Pasadena, CA 91109, USA}

\author[0000-0002-8452-8675]{J.\ Bernard-Salas}
\affiliation{ACRI-ST, Centre d’Etudes et de Recherche de Grasse (CERGA),
  10 Av.\ Nicolas Copernic, 06130 Grasse, France}
\affiliation{INCLASS Common Laboratory, 10 Av.\ Nicolas Copernic, 06130
  Grasse, France}

\author[0000-0002-1693-2721]{D. A. Garc\'{\i}a-Hern\'andez}
\affiliation{Instituto de Astrof\'{\i}sica de Canarias, C/ Via L\'actea
s/n, E-38205 La Laguna, Spain}
\affiliation{Departamento de Astrof\'{\i}sica, Universidad de La Laguna
(ULL), E-38206 La Laguna, Spain}

\author[0000-0003-0665-6505]{J.\ Li}
\affiliation{Instituto de Astrof\'{\i}sica de Canarias, C/ Via L\'actea
  s/n, E-38205 La Laguna, Spain}
\affiliation{Departamento de Astrof\'{\i}sica, Universidad de La Laguna
  (ULL), E-38206 La Laguna, Spain}

\author[0009-0000-0191-6756]{C.\ Bhatt}
\affiliation{Department of Physics and Astronomy, The University of Western Ontario, London, ON N6A 3K7, Canada}
\affiliation{The Institute for Earth and Space Exploration, The University of Western Ontario, London, ON N6A 3K7, Canada}

\correspondingauthor{A.~Ricca}
\email{Alessandra.Ricca-1@nasa.gov}

\begin{abstract}
The infrared spectra of many carbon-rich post-asymptotic giant branch stars are dominated by emission features from aromatic and aliphatic hydrocarbons, but the chemical structure of the carriers remains unidentified. Class D sources show unusual emission profiles with strong aliphatic emission, providing a stringent test of carrier candidates. 
Here we investigate whether hydrogenated carbonaceous molecular particles can be the carriers of these features. Rather than assuming candidate geometries we generate the initial structures using ab initio molecular dynamics simulations and optimize them using density functional theory. We then compare the computed emission spectra to JWST NIRSpec and MIRI/MRS observations of Class D sources. The structures providing the best agreements contain sizeable domains with defect-bearing aromatics and hydrogenated-aromatics connected by aliphatic bridges, in contrast to the ``arophatic" cluster model of isolated two- to three-ring aromatics connected by aliphatic and olefinic bridges. 
As a representative example, we discuss in detail the molecular particle C$_{130}$H$_{96}$ (radius of 5.5~\AA) and compare its computed spectrum to the JWST spectra of IRAS 05110-6616, a Class D2 source with apparent shifts of the aromatic C–-H out-of-plane bands. The calculated spectra are in qualitative agreement with the observed 3~\micron\ profile, the aliphatic C-–H bending bands, the broad 8~\micron\ feature, the 11–-14~\micron\ C–-H out-of-plane region, and the bands at 16 and 21~\micron.
We hypothesize that UV photo-driven fragmentation of such molecular particles releases aromatics linked by aliphatics, followed by the formation of polycyclic aromatic hydrocarbons and fullerenes as these objects evolve into planetary nebulae.

\end{abstract}


\section{Introduction}
\label{sec:introduction}
Carbon-rich stars are important sources of carbonaceous materials in the ISM. After ejection, the material is further processed by UV radiation, energetic particles and strong shocks and ends with its incorporation into newly formed stars and planetary systems. Understanding the currently unknown steps along this path and elucidating the chemical structures of the emission carriers is of key importance. It is generally thought that photo-processing can change the ratio of aliphatic carbon (present in chains or rings, e.g., sp$^3$ hybridized C containing four single bonds in a tetrahedral geometry) versus aromatic carbon (present in aromatic rings, e.g., sp$^2$ hybridized C containing one double bond and two single bonds in a trigonal planar geometry) \citep{Jochims_1994,Sloan_1997,Jochims_1999,Tielens_2008,Peeters_2024}.

In some benign environments associated with post-asymptotic giant branch (post--AGB) stars and protoplanetary nebulae (PPNe), a relatively high aliphatic fraction has been observed in infrared (IR) emission spectra. Aliphatic bands peak at 3.4--3.5~\micron\ \citep{Geballe_Van_der_Veen_1990,Geballe_1992,Geballe_1994} and at 6.9 and 7.25~\micron\ in addition to broad features near 8~\micron\ and in the 11--15~\micron\ region \citep{Sloan_2007,Sloan_2014,Sloan_2017,Matsuura_2014,Jensen_2022}. 
In harsh environments associated with H\textsc{ii} regions, Herbig AeBe stars, reflection nebulae (RNe), diffuse ISM, and planetary nebulae (PNe), the fraction of C atoms in aliphatic form observed in the IR emission spectra is much lower \citep[10\%; e.g.,][]{Li_Draine_2012,Yang_2016}. These spectra are classified as Class A and B \citep{Peeters_2002}. In these objects, aromatic bands at 6.2, 7.7, 8.6, 11.2 and 12.7~\micron, generally attributed to polycyclic aromatic hydrocarbons (PAHs), dominate the spectra, while aliphatic bands at 6.9 and 7.25~\micron\ are negligible. Class C sources include Young Stellar Objects \citep[YSOs;][]{Acke_2010} and post--AGB stars \citep{Peeters_2002,Sloan_2007}. Their emission spectra contain broader features than Class A and B spectra and the aliphatic/aromatic content is in-between that of Classes A/B and D sources \citep{Sloan_2007,Pino_2008}.
Class D sources have been further divided as Class D1 and D2 \citep{Matsuura_2014,Sloan_2014}. Their emission carriers, which we will refer here as ProtoPAHs, have spectra that are inconsistent with PAH spectra \citep[in press]{Clark_2026}. The chemical structure of the carriers remains elusive.

\textit{JWST} program ID 4678 (PI: G.~C. Sloan) observed a sample of seven targets, all of them carbon-rich post-AGB objects in the Large Magellanic Cloud (LMC) and the Small Magellanic Cloud (SMC). The sample includes two sources with Class D1 PAH spectra, two D2, two sources known or suspected
to have strong absorption spectra, and one Class B source complemented with SMP LMC 058, a PN with Class B PAH emission
that was observed as a calibration target by \textit{JWST}. Class D observed spectra provide the observational constraints needed to characterize ProtoPAHs ~\citep[submitted]{Sloan_2026}.

The infrared emission of mixed aromatic/aliphatic carriers has previously been modeled with density functional theory, using hand-constructed structures ranging from PAHs with aliphatic side groups \citep{Sadjadi_2015,Sadjadi_2017} to mixed aromatic/aliphatic organic nanoparticles \citep[MAONs;][]{Kwok_Zhang_2011,Kwok_2013,Sadjadi_2016}. 

\citet{Jones_Ysard_2025} used THEMIS \citep[The Heterogeneous dust Evolution Model for Interstellar Solid;][]{Jones_2013} to generate a three-dimensional ``arophatic" cluster model, consisting of isolated clusters of two to three six-membered ring aromatics connected by aliphatic and olefinic bridges \citep{Jones_2012c,Micelotta_2012}. They further assessed the photon-driven fragmentation and Coulomb fragmentation of the ``arophatic" cluster model. Alternatively, the molecular particles can be computed by starting from its precursor material, namely amorphous carbon (a-C). Computational ab initio molecular dynamics (AIMD) methods provide realistic models of a-C for densities spanning 0.95 to 3.5 g cm$^{-3}$ \citep{Bhattarai_2018,Bauschli_Lawson_2010}. The structures vary from interconnected wrapped and defective sp$^2$ sheets at 0.95 g cm$^{-3}$ to almost tetrahedral (diamond--like) structures at 3.5 g cm$^{-3}$ \citep{Bilek_2000}.

In this work we seek to characterize the ProtoPAH emitters without prior assumptions about their structure by using a-C as a starting point and the constraints provided by \textit{JWST} observations. We should note that these structures are not being proposed as carriers of the general AIB spectrum, but only as carriers of the specific sources that show a spectrum that is markedly different from the regular AIBs, such as the proto-PAH targets.

This paper is organized as follows. Section \ref{sec:methods} describes the theoretical methods employed in this paper. Section \ref{sec:results} provides an analysis of the infrared (IR) spectral characteristics and structural properties of the molecular particle. Further astrophysical implications for these findings are discussed in Section \ref{sec:discussion}, and Section \ref{sec:conclusions} summarizes these results.

\section{Methods}
\label{sec:methods}

\begin{figure*}
    \centering
    \includegraphics[width=\textwidth]{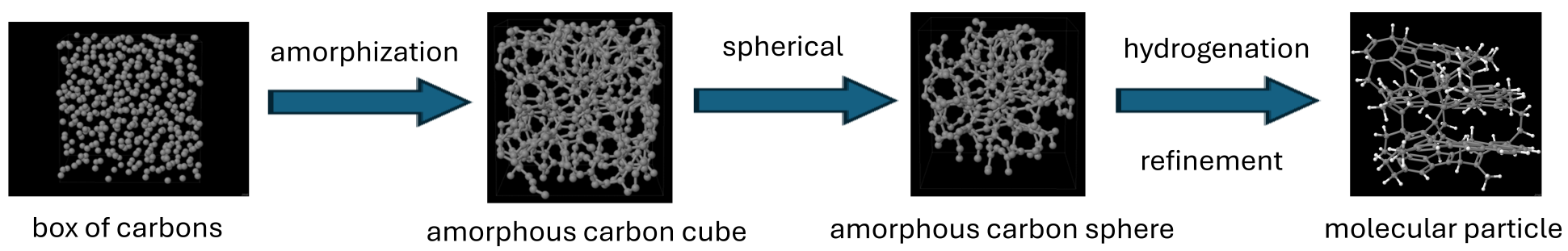}
     \caption{Steps involved in generating the hydrogenated carbonaceous molecular particle structures.
     \label{fig:process}}
\end{figure*}

The steps involved in generating the hydrogenated carbonaceous molecular particle structures, with a net charge of zero, are shown in Figure~\ref{fig:process}. Starting from a cubic box containing carbon atoms placed randomly we obtained bulk a-C. We generated a sphere out of the amorphous carbon cube and then hydrogenated it. Linear chains containing carbon atoms doubly coordinated (C$_{sp}$) were removed and long aliphatic chains were folded to allow ring formation. The geometry of the sphere was subsequently optimized. 
 
Bulk a-C was computed using ab initio molecular dynamics (AIMD) simulations and employing the Vienna ab initio simulation package (VASP), version 5.4.4.18Apr17 \citep{Kresse_1993,Kresse_1994,Kresse_1996a,Kresse_1996b}. The Perdew-Wang 91 (PW91) exchange-correlation functional \citep{Perdew_1992} and a plane wave basis set, in conjunction with the projector augmented wave method \citep[PAW;][]{Kresse_1999}, were employed. An energy cutoff of 400~eV for the plane-wave basis set and a $1 \times 1 \times 1$ grid of \textit{k} points were used in all simulations.

We generated a cubic cell containing 128 carbon atoms with a density of 2.0 g cm$^{-3}$. The atoms were distributed randomly in the box with the constraint that their positions be at least 0.5~\AA\ away from the edges and that the distance between the carbon atoms be greater that 1.3 \AA. The cubic cell was then replicated along the x,y,z directions to produce a bulk structure. The atomic positions in the cell were relaxed at 0~K while keeping the dimensions of the cell fixed.

The amorphization step, shown in Figure~\ref{fig:process}, was performed following the procedure described by \citet{Bauschli_Lawson_2010} using three steps: (\textit{i}) melting consisting in heating and equilibrating at 5000~K over 0.6 ps; (\textit{ii}) quenching from 5000~K to 1~K over 4 ps and (\textit{iii}) energy minimization at 0~K.

From the resulting a-C structure we extracted a sphere with a radius of 7~\AA\ that captured enough aliphatic, aromatic, and olefinic (non-aromatic C=C double bonds) components. The bond connectivity was assessed and defects (carbon atoms with less than four bonds) were hydrogenated.

The geometries of the neutral molecular particles obtained from the MD simulations were reoptimized using the density functional theory (DFT) method, the 6-31G* basis set \citep{frisch84} and their harmonic frequencies computed using the Gaussian~16 suite of programs \citep{Gaussian16}. We ensured that the structures had a minimum on the potential energy surface, i.e. they didn't produce imaginary frequencies.

The harmonic frequencies were scaled to lower values using three scaling factors, namely 0.959 for C-H stretches in the 3~\micron\ region, 0.9715 for the 5-10~\micron\ region, and 0.9821 for everything above 10~\micron\ \citep[][]{bauschli2018pahdb}. The frequencies (in cm$^{\rm -1}$) and band intensities (in km mol$^{\rm -1}$) were synthesized into emission spectra using an emission cascade model. Emission spectra were computed for the absorption of a single photon using excitation in the solar radiation field characterized by a 5770~K blackbody ($\approx$ 3.5 eV), and the entire cooling cascade was taken into account \citep[see e.g.,][]{Boersma2010, Boersma2011, Boersma2014, Bauschli_2010}. The integration was done over wavenumbers (units of erg) and convolution with a line profile gave units of radiant energy in erg/cm$^{-1}$. Gaussian line profiles were used with full width at half maximum (FWHM) values of 30~cm$^{\rm -1}$ for bands shortward of 9~\micron\ and 25~cm$^{\rm -1}$ for bands between 9--20~\micron, based on comparisons with observations.

The geometries and vibrational normal modes were visualized using the software JMol\footnote{\url{https://jmol.sourceforge.net/}}. All the structural and spectroscopical data will be made publicly available in the NGdb -- Theoretical database\footnote{\url{https://nanograin.odr.io/}}.

This work is based on observations made with the NASA/ESA/CSA James Webb Space Telescope. All of the data presented in this article were obtained from the Mikulski Archive for Space Telescopes (MAST) at the Space Telescope Science Institute. These observations are associated with program \#4678 (DOI: 10.17909/zs5n-2t21).

\section{Results}
\label{sec:results}

\subsection{Structural properties}
\label{subsec:morphology}
\begin{figure}[h]
   \includegraphics[width=1.0\linewidth]{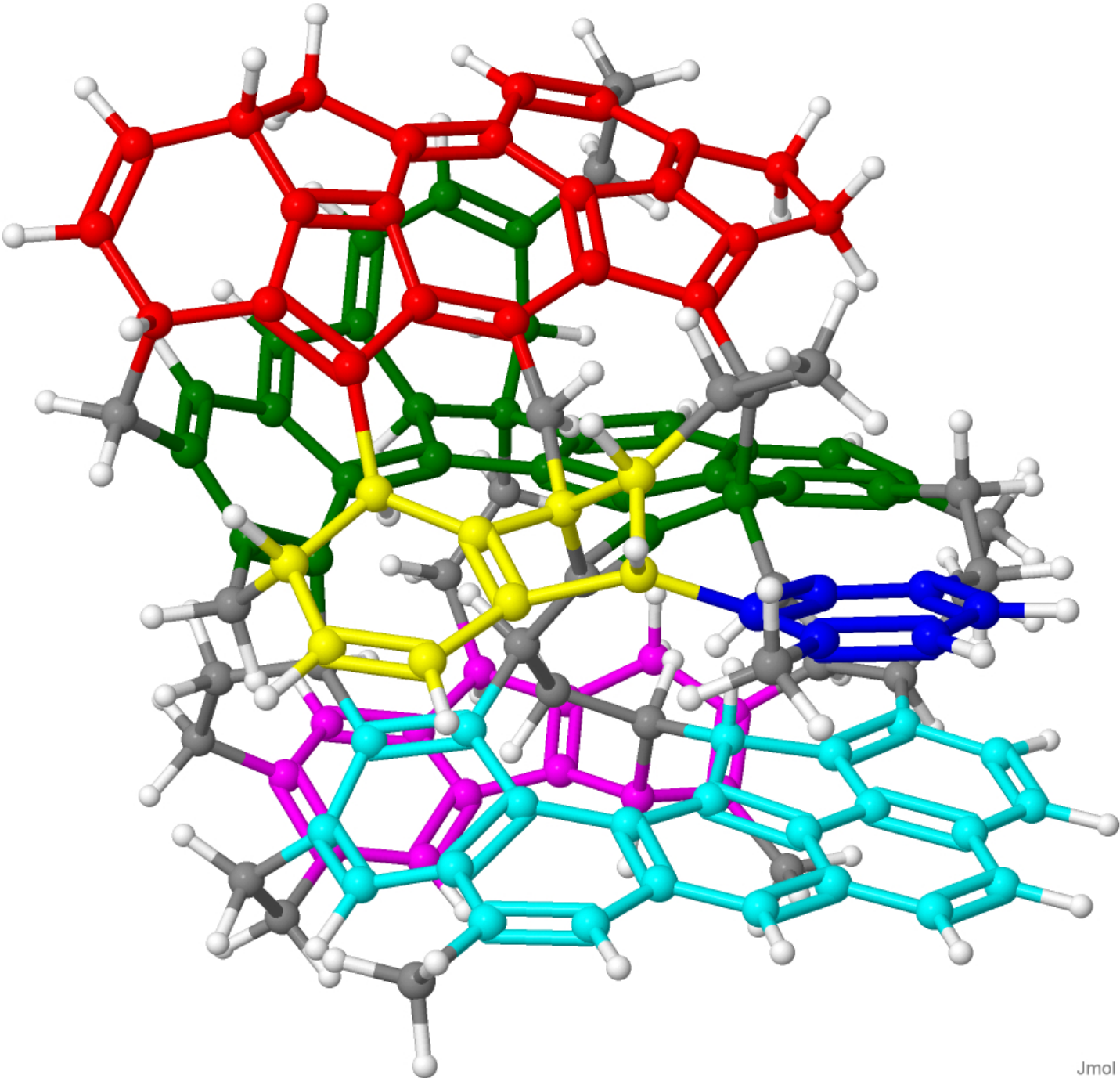}
    \caption{Structure of the hydrogenated carbonaceous molecular particle C$_{130}$H$_{96}$ with the various aromatic/hydrogenated aromatic domains, highlighted in different colors, connected by aliphatic links, shown in grey.}
    \label{fig:structure}
\end{figure}

We obtained seventy structures and assessed their relevance based 
on visual comparisons of their modeled emission spectra with \textit{JWST} observations. Poor agreements were obtained when (\textit{i}) olefinic double bond chains linked aromatics and hydrogenated aromatics (i.e. aromatics with additional hydrogen atoms) and (\textit{ii}) when only a very small number of aromatic rings were present in the domains.
The best agreements were obtained for molecular particles containing sizeable aromatic/hydrogenated aromatic domains (bright colors) connected by aliphatic links (grey color), as shown in Figure~\ref{fig:structure}. A molecular particle representative of this class is C$_{130}$H$_{96}$ with a radius of 5.5~\AA. C$_{130}$H$_{96}$ itself is not a unique astronomical carrier. Rather, it serves as a representative example demonstrating that a single molecular particle containing aromatics and aliphatics can account for several features that
have previously been difficult to explain simultaneously, and that such a structure can indeed reproduce many of the observed characteristics. This particular structure with its curved and defective ring systems, hydrogenated aromatic regions, and aliphatic connections may provide clues to how such particles could later fragment or rearrange into PAHs and fullerenes. In a subsequent paper we plan to extend this study to an ensemble of structures and include distributions in size, H/C, and C$_{sp^2}$/C$_{sp^3}$ ratios.

For each domain, the breakdown of aromatic and hydrogenated aromatic rings, the total number of connections to the rings and the number and type of aliphatic links are given in Table~\ref{tab:domains}. Most of the fully aromatic rings are six--membered rings whereas the hydrogenated aromatic rings contain five--, six-- and seven--membered rings. Most of the aliphatic C--H links contain up to two carbons, with a few links containing three and four carbons.
The number of aromatic carbons is 67, olefinic carbons is 16 and aliphatic carbons is 47. Twelve aliphatic hydrogens are in 4 CH$_3$ groups, 46 aliphatic hydrogens are in 23 CH$_2$ groups, and 16 aliphatic hydrogens are in 16 CH groups. Together with the 15 aromatic and 7 olefinic H atoms, this gives a total of 96 H atoms. This mixed character is due to the presence of defects, such as pairs of five/seven--membered rings, which reduce the number of aromatic bonds and create carbon radicals (with less than four bonds) which can react with hydrogen atoms to form hydrogenated aromatic rings and/or with other carbon radicals to form C--C bonds. 
The presence of five--membered rings induces curvatures in the structure that depend on the number of five--membered rings and their relative positions. The rings are linked by aliphatics attached at their edges.
The structure is more strained and disordered than PAHs linked by aliphatic chains. It can be viewed as an intermediate component between a-C and PAHs.

\begin{deluxetable*}{lcccccccccccccc}[h]
\tiny
    \tablecaption{Breakdown of the number of aromatic and hydrogenated aromatic rings, along with the total number of carbon links attached to each domain (total C-links), and the type and number of aliphatic links (CH$_3$, CH$_2$, 2(CH$_2$) (2C), CHCHCH$_3$ (3C-1), CHCH$_2$CH (3C-2), CHCH$_2$CH$_2$ (3C-3), CH$_2$CHCH$_2$CH$_2$ (4C)).\label{tab:domains}}
    \tablehead{
		\colhead{} & \multicolumn{3}{c}{Arom. rings} &  \multicolumn{3}{c}{Hydr. arom. rings} &\multicolumn{1}{c}{Total C-links} & \multicolumn{7}{c}{Aliphatic links}\\
	\colhead{Domain} & \colhead{Five}& \colhead{Six}& \colhead{Seven}& \colhead{Five}& \colhead{Six}& \colhead{Seven}&  & \colhead{CH$_3$}& \colhead{CH$_2$}& \colhead{2C}& \colhead{3C-1}& \colhead{3C-2}& \colhead{3C-3}& \colhead{4C}}
	\startdata
	Red&       2&  1&  1&  2&  1&  1&  5&   &   2&   1&    1\\
    Green&      &  5&   &  1&   &  1&  8&   &   2&   4&    1\\
    Magenta&    &  1&   &  1&  1&   &  6&  1&    &   3&    &   1&  &  1\\
    Cyan&       &  6&   &  1&   &   &  6&  1&    &   1&    &   1&  1&  1\\
    Blue&       &  1&   &   &   &   &  4&  1&   1&   1&   \\
    Yellow&     &   &   &  1&  1&   &  6&   &    &    &   \\
	\tableline
	\enddata
\end{deluxetable*}


\begin{table*}
  \centering
\caption{Summary of computed band positions (\micron), intensities (kcal mol$^{-1}$), and assignments for C$_{130}$H$_{96}$.}\label{table:emission_sum}
\begin{tabular}{lll}
\hline \hline
Band position & Intensity & Assignment \\
\hline
3.278 &  0.023737 &  aromatic C--H stretching   \\
3.424 &  0.087557 &  CH$_2$ asymmetric stretching in aliphatic links and CH$_3$ asymmetric stretching \\  
3.50  &  0.02158  &  CH$_2$ symmetric stretching in hydrogenated aromatics and CH$_3$ symmetric stretching \\
6.261 &  0.797    &  C--C aromatic stretching \\
6.623 &  1.10     &  C-H in-plane bending and CH$_2$ scissoring in long aliphatic links\\
6.813 &  1.72     &  CH$_2$ in-plane scissoring in short aliphatic links \\
7.196 (blended) &  --       &  CH$_3$ symmetric bending (umbrella mode)  \\
7.372 -- 9.25 &  9.78  &  aliph., arom., olef. in-plane C-H bending, skeletal C--C distorsions \\
10.961  &  3.99  &  aromatic CH$_{oop}$ solo and CH$_2$ in-plane rocking \\
11.242  &  5.34  &  aromatic CH$_{oop}$ solo \\
11.783  &  2.03  &  aromatic CH$_{oop}$ solo and CH$_2$ in-plane rocking  \\
12.444  &  3.82  &  aromatic CH$_{oop}$ duo \\
13.371  &  2.98  &  CH$_2$ in-plane rocking and CH$_{oop}$ solo \\
16.16   &  3.70 &  skeletal bending\\
18.63   &  2.79  &  skeletal bending \\
20.70   &  2.22  &  skeletal bending\\
\hline \hline
\end{tabular}
\end{table*}

\begin{figure*}
\centering
\subfigure{
     \includegraphics[scale=.3]{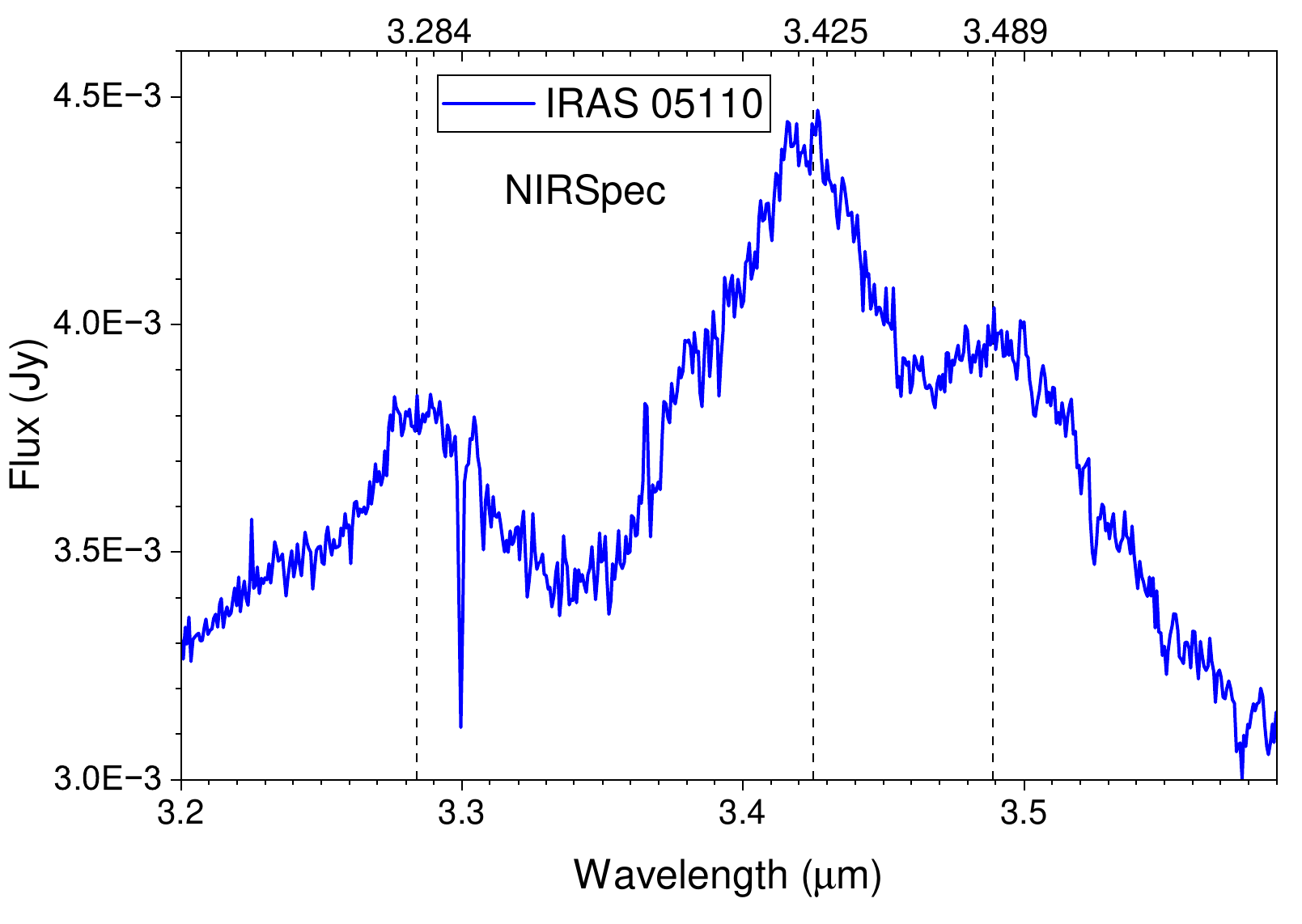}
      \label{fig:nrs}
}
\hfil
\subfigure{
     \includegraphics[scale=.3]{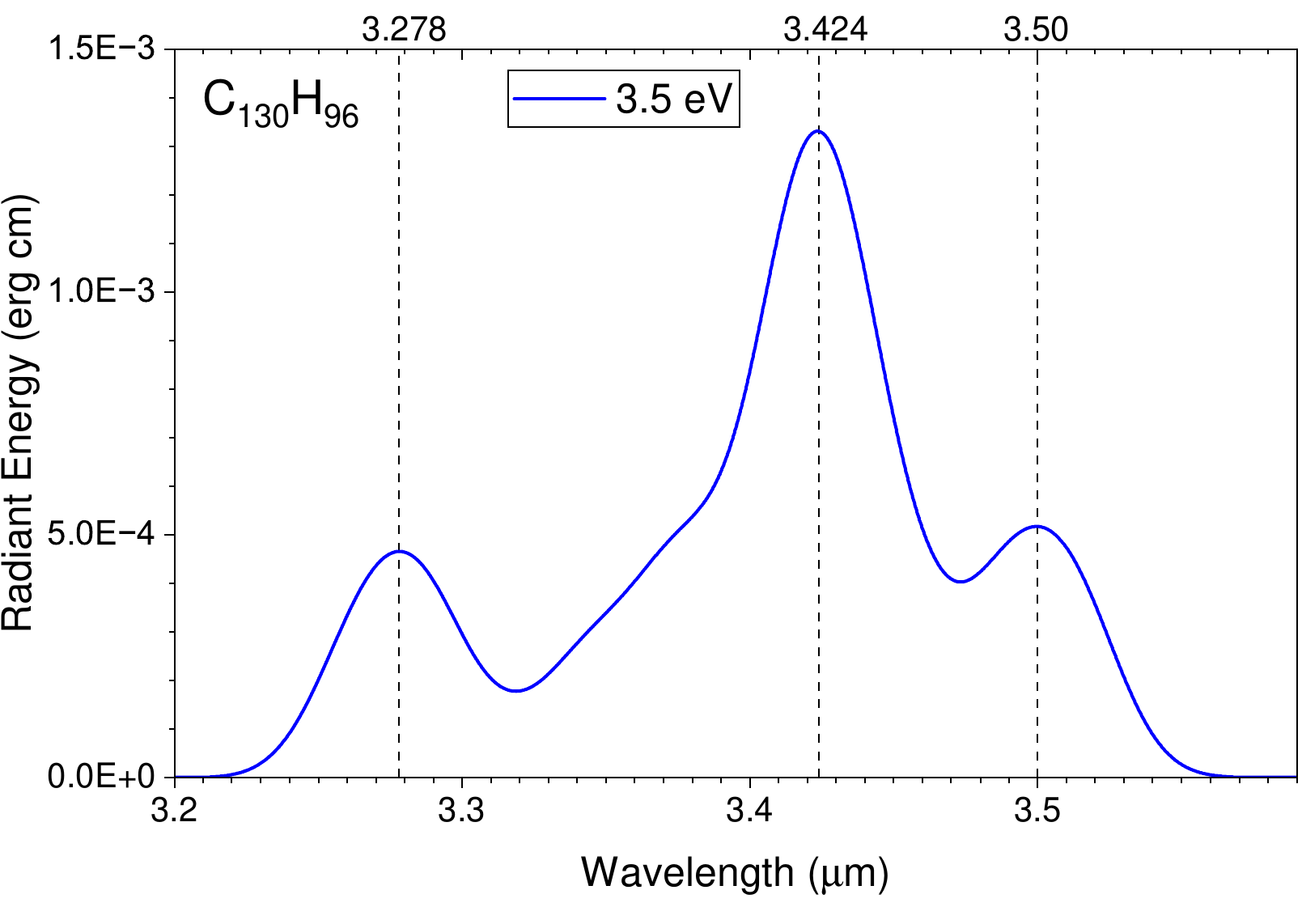}
      \label{fig:c130_3mu}        
}
\hfil
\subfigure{
     \includegraphics[scale=.3]{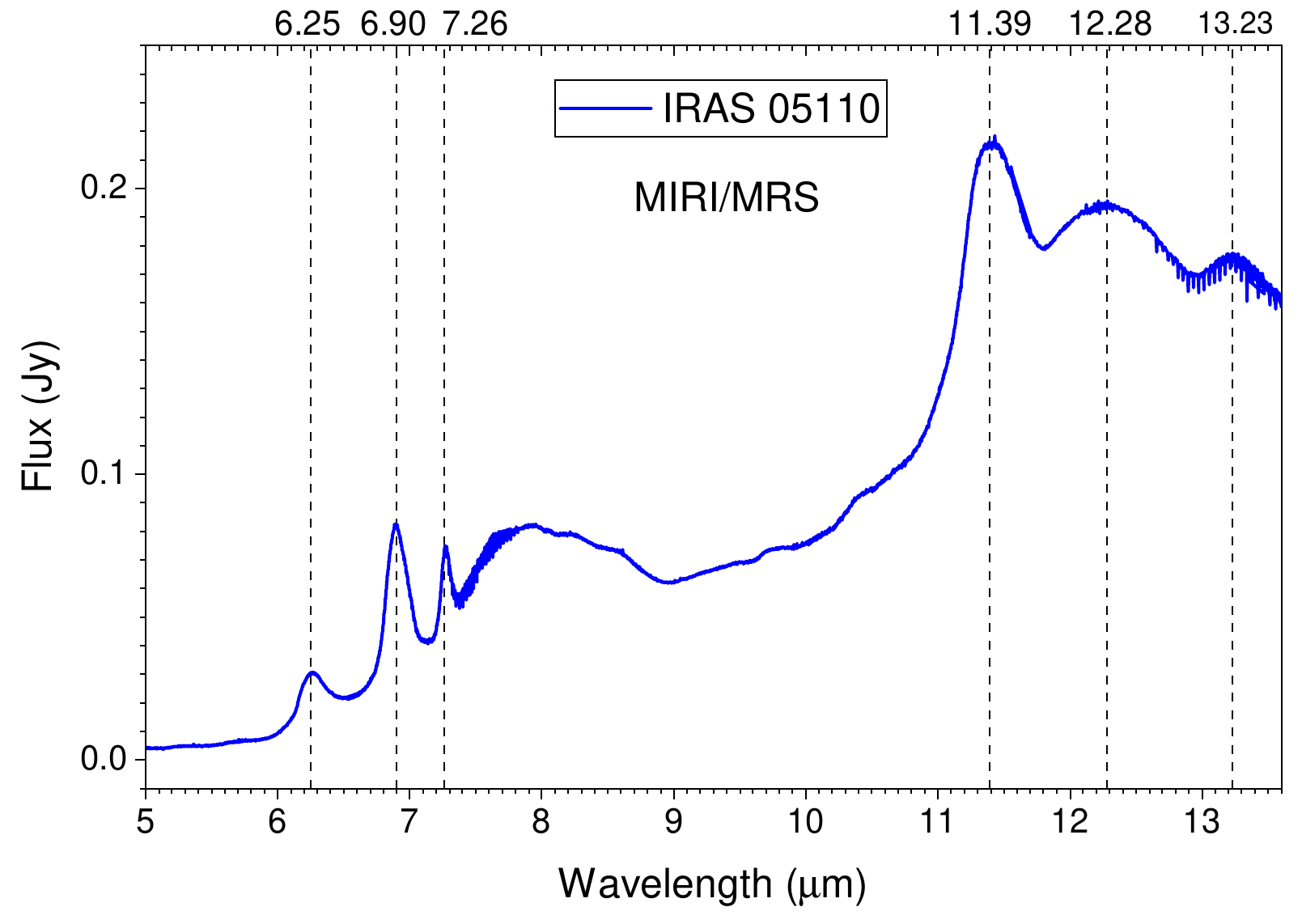}
      \label{fig:miri_5-14}        
}  
\hfil
\subfigure{
     \includegraphics[scale=.3]{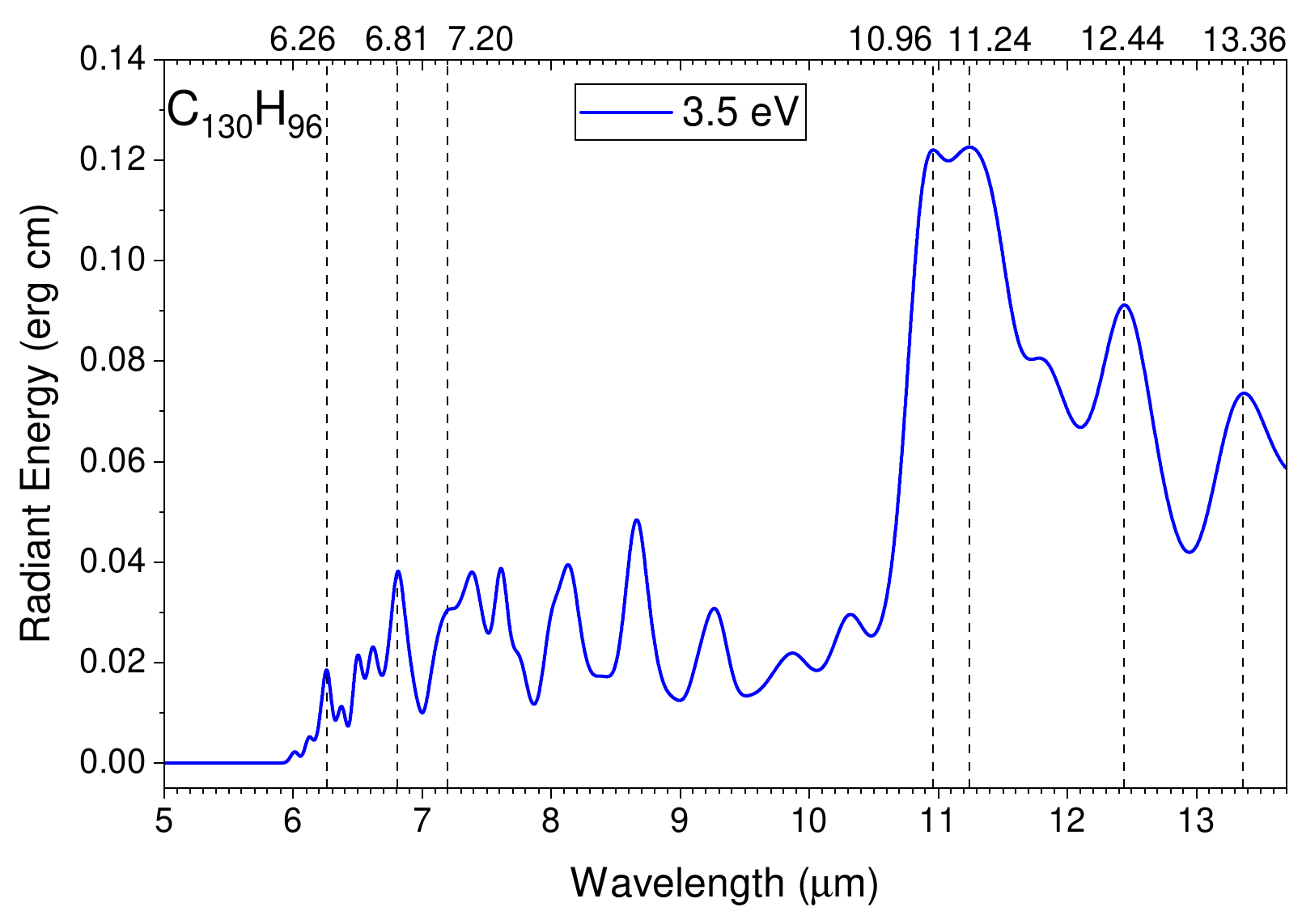}
      \label{fig:c130_5-13_7mu}
}
\hfil
\subfigure{
     \includegraphics[scale=.3]{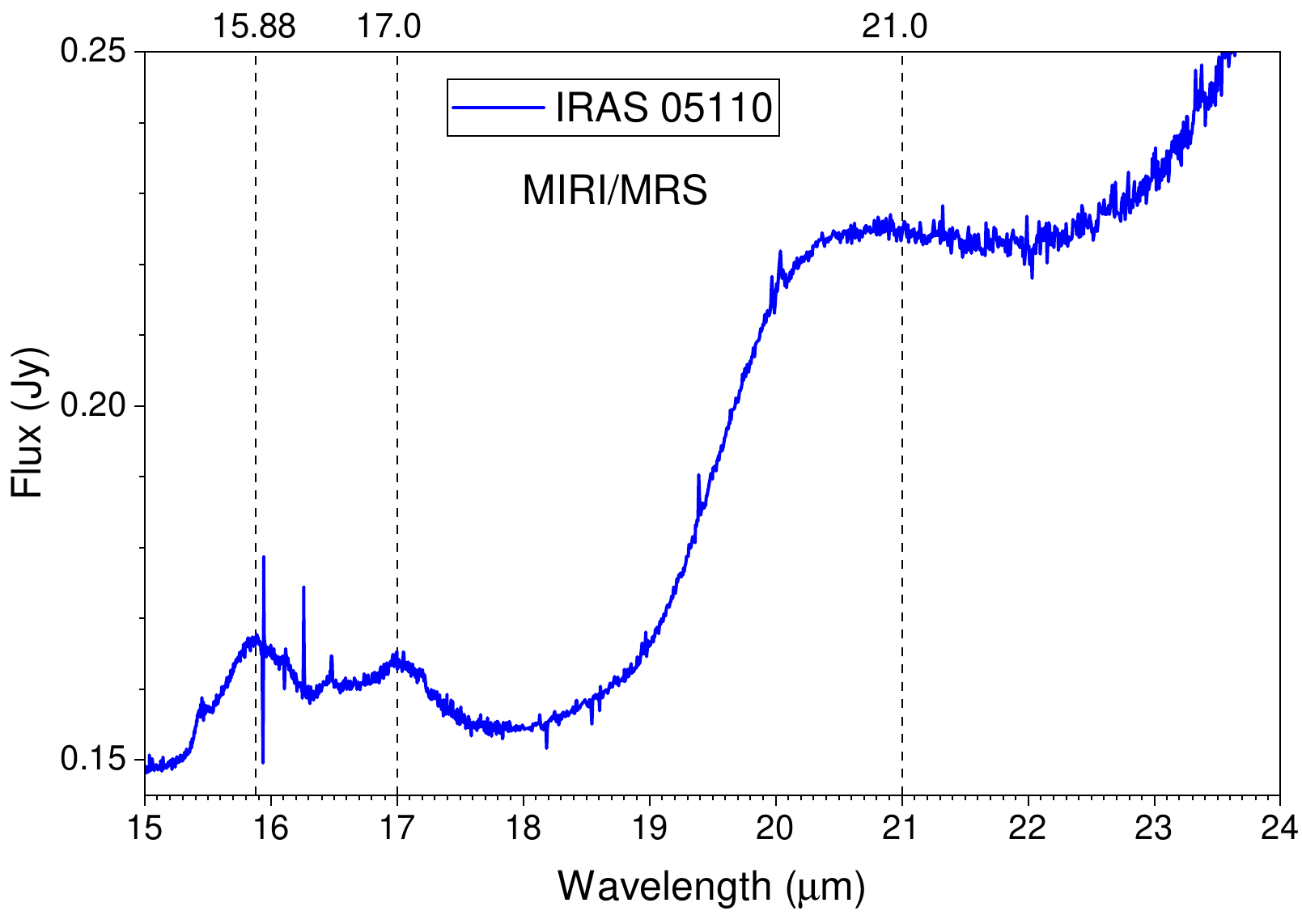}
      \label{fig:miri_15-24}        
}
\hfil
\subfigure{
     \includegraphics[scale=.3]{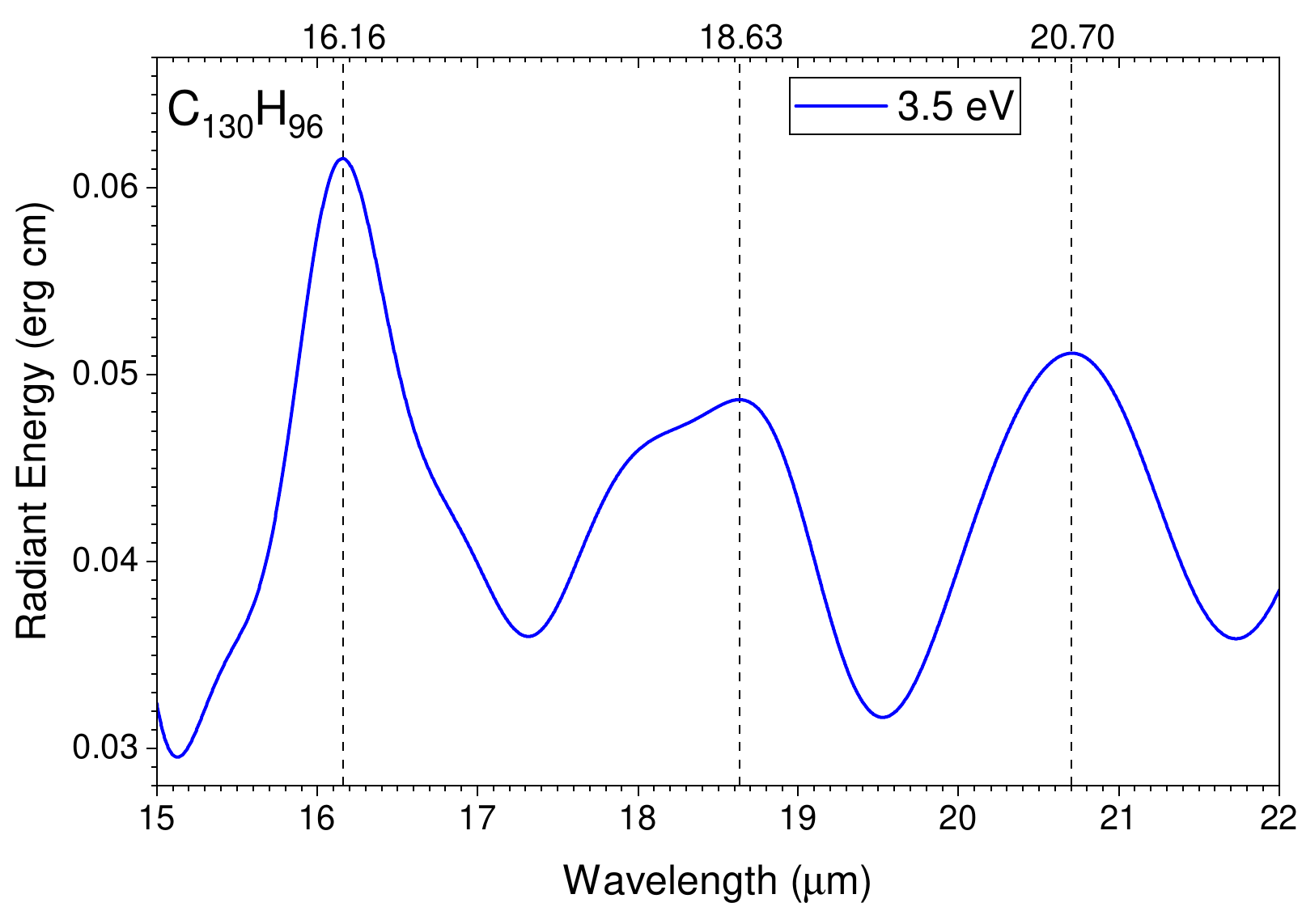}
      \label{fig:c130_15-22mu}
}
\caption{Comparison of \textit{JWST} spectra of IRAS 05110 (left) with computed emission spectra for the absorption of a 3.5~eV photon (right). Dotted vertical lines indicate band positions.}
\label{fig:emission}
\end{figure*}

\subsection{Emission spectra}
\label{subsec:emission}
NIRSpec and MIRI/MRS spectra of the Class D2 IRAS 05110-6616 (IRAS 05110 hereafter) source \citep[submitted]{Sloan_2026} are shown in Figure~\ref{fig:emission} along with the computed emission spectra of the neutral molecular particle C$_{130}$H$_{96}$, for the absorption of a 3.5~eV photon. The computed band positions, intensities, and assignments are given in Table~\ref{table:emission_sum}.

A detailed analysis of the NIRSpec spectrum of IRAS 05110 was performed using gaussian fits of the normalized, continuum-subtracted emission bands \citep[in press]{Clark_2026}. It showed that the main emission bands in the NIRSpec range can be decomposed into three features: \textit{i)} a wide aromatic C--H stretching band at 3.284~\micron\ with some olefinic C--H stretching contributions at 3.24 and 3.35~\micron; \textit{ii)} a CH$_2$ asymmetric stretching band at 3.425~\micron\ with a shoulder at 3.38~\micron\ due to a CH$_3$ asymmetric stretching band; \textit{iii)} an aliphatic band at 3.489~\micron\ with contributions from a CH$_3$ symmetric stretching band (3.4838~\micron) and a CH$_2$ symmetric stretching band (3.5075~\micron).
No strong band was observed at 3.40~\micron\ which indicates that superhydrogenated aromatics, such as 1,2,3,6,7,8-hexahydropyrene \citep{Demyk_2026}, are not present in substantial amounts.

The computed 3~\micron\ emission (Figure~\ref{fig:emission})
contains three bands that fall at wavelengths comparable to observations: \textit{i)} an aromatic C--H stretching band at 3.278~\micron\ with a small olefinic C--H stretching band contribution at 3.267~\micron; \textit{ii)} a CH$_2$ asymmetric stretching band in aliphatic links at 3.424~\micron\ with an inflection point at 3.38~\micron\ due to a CH$_3$ asymmetric stretching mode and a blue rise starting at 3.32~\micron\ due to olefinic C--H and CH$_2$ stretching modes; \textit{iii)} a band at 3.50~\micron\ due to a CH$_2$ symmetric stretching mode in hydrogenated aromatics and a CH$_3$ symmetric stretching mode. Our spectra do not produce a strong peak at 3.40~\micron\ consistent with the absence of superhydrogenated aromatic domains.
The ratio 3.4 aliphatic/$\sum(3.3, 3.4)$ is 0.82, which is slightly larger than the value of 0.7749 from IRAS 05110 \citep[in press]{Clark_2026}. The bands at 3.278 and 3.50~\micron\ have comparable intensities and are a factor of 3.9 smaller than the 3.424~\micron\ band. The larger intensity of the 3.424~\micron\ band is consistent with the structure containing more CH$_2$ groups in aliphatic links than in hydrogenated aromatics. Despite a larger CH$_2$/CH$_3$ ratio than observed for IRAS 05110 the overall computed 3~\micron\ profile is in qualitative agreement with the \textit{JWST} profile. We should note that large structures with 150 carbon atoms or more are expected to produce very weak 3~\micron\ bands for the absorption of a 3.5~eV because the absorbed energy is distributed over more vibrational modes in a larger particle. 

The MIRI/MRS spectrum contains several distinct features: \textit{i)} an aromatic C--C stretching band at 6.25~\micron; \textit{ii)} a CH$_2$ asymmetric bending band and a CH$_3$ asymmetric bending band at 6.90~\micron; \textit{iii)} a CH$_3$ symmetric bending band at 7.26~\micron; \textit{iv)} a broad feature extending from $\approx$ 7.4 to 9~\micron\ of unknown origin; \textit{v)} three C--H out-of-plane bending (OOP) bands at 11.39, 12.28 and 13.23~\micron\ with decreasing intensities relative to each other \citep[in press]{Clark_2026}.
In addition, in the far-IR the MIRI/MRS spectrum contains bands at 15.88, 17.0 and the ``21.0~\micron" feature \citep[submitted]{Sloan_2026}.

The computed 5.0--13.7~\micron\ emission (Figure~\ref{fig:emission}) shows that the intensities of the bands in the 6--10~\micron\ are smaller than those of OOP bands (11--14~\micron\ region) by approximately a factor of 3, consistent with neutral molecules \citep{Langhoff_1996}.
The aromatic C--C stretching peaks at 6.26~\micron\, in agreement with observations, and a weak olefinic C--C stretching band is visible at 6.12~\micron. The position of the CH$_2$ scissoring mode is dependent on the level of molecular strain. In short aliphatic links, it peaks at 6.811~\micron\ and in hydrogenated aromatic rings it peaks around 6.9~\micron\ \citep{Sandford_2013,Materese_2017}. For long aliphatic links with low strain the scissoring mode blueshifts to 6.67~\micron. Table~\ref{tab:domains} shows that our structure contains more CH$_2$ groups in short aliphatic links than in hydrogenated aromatics, which is reflected in the CH$_2$ scissoring mode peaking at 6.81~\micron. A few weaker CH$_2$ scissoring modes at 6.6~\micron\ are due to aliphatic links containing three and four carbons. In observations, the majority of emission is centered at 6.90~\micron, with only a small contribution at 6.81~\micron, and has been attributed to an aliphatic C–H scissoring mode adjacent to an aromatic ring \citep[in press]{Clark_2026}. As the position of the 6.9~\micron\ band is sensitive to the specific chemical structure it is possible that the discrepancy between our calculations and observations is due to the specific structure of our molecular particle. The computed CH$_3$ symmetric bending mode (i.e. umbrella mode) peaks at 7.20~\micron, at a slightly shorter wavelength than the observed band, and it is followed by a series of peaks attributed to combinations of aliphatic, aromatic, and olefinic C--H bending modes that are coupled with the collective skeletal C--C modes. Given the disordered and distorted nature of these molecular particles and the fact that multiple structures are likely to be present, it is expected that many bands from different species will overlap and form a broad feature in this range \citep{Lazzarini_2016} as seen in the IRAS 05110 MIRI/MRS spectrum.

Two weak computed olefinic OOP modes peak at 10.34 and 10.85~\micron\ and a weak CH$_2$ in--plane rocking band peaks at 9.87~\micron.
The ``11.2~\micron" feature is broad and consists of a band at 10.96~\micron, attributed to a combination of an aromatic C--H solo OOP mode and an aliphatic CH$_2$ in-plane rocking mode (in aliphatic link and hydrogenated aromatic), and a band at 11.24~\micron\ assigned to an aromatic C--H solo OOP band. This broadening is consistent with the modification of OOP bands by the presence of aliphatics \citep{Sandford_2013, Dartois_2020}. The band at 12.44~\micron\ is attributed to an aromatic C--H duo OOP mode, and the band at 13.36~\micron\ is due to a combination of an aliphatic CH$_2$ in-plane rocking and an aromatic C--H solo OOP mode. The interaction of aromatics and aliphatics explains the shifting of the bands observed in the Class D2 spectra.

In the far-IR we obtained three bands at 16.16, 18.63, and 20.70~\micron\ due to skeletal bending modes. The computed band positions at 16.16 and 20.70~\micron\ are in agreement with observations whereas the band at 18.63~\micron\ is redshifted. The relative intensity of the 20.70~\micron\ band is significantly different from observations. Discrepancies in the far-IR region are typically due to the fact that the computed spectra in this region are very sensitive to the overall structure. Slight differences in structure can produce entirely different skeletal modes. 
We should note that other carbonaceous particles are likely to contribute to the 21~\micron\ band \citep[see e.g.,][]{Papoular_2013,Volk_2020} and possibly explain the strong intensity of the 21~\micron\ band. We should stress that our molecular particle is another example of a possible candidate.

Overall, our computed spectra show all the general features observed for IRAS 05110 and one can envision that an ensemble of hydrogenated carbonaceous molecular particles could provide an even better agreement.

\section{Discussion}
\label{sec:discussion}
Our computed molecular particle is a representative example demonstrating that a single molecular particle containing aromatics and aliphatics can account for several features that have previously been difficult to explain simultaneously and reproduce many of the observed emission characteristics. It supports the hypothesis by \citet[in press]{Clark_2026} that the emission, in a small subset of Class D carriers, is due to a single component instead of two coexisting components. They based their hypothesis on the position of the 6.9~\micron\ feature and the perturbation of the aromatic OOP modes which suggested that the aromatic and aliphatic groups are close to one another within the same carrier.
The computed structure of our representative component contains aromatic/hydrogenated aromatic domains, with pairs of five--/seven--membered ring defects, connected by aliphatic links. It is different from the three-dimensional ``arophatic" cluster model, which is less strained and contains only small defect-free aromatic domains without the presence of hydrogenated aromatics. 

During the transition from the post-AGB phase to PNe, the molecular particles will experience harsher conditions and UV photon-driven fragmentation. \citet{Quan-De_Wang_2017} studied the thermal decomposition of n-hexene (a partially hydrogenated benzene) and n-hexane (a fully hydrogenated benzene). The resulting bond (marked by a $\wr$) energies were 3.2~eV for an allyl-adjacent group, --C-$\wr$-C--C=C--, present in n-hexene, 3.8~eV for an aliphatic --C-$\wr$-C-$\wr$-C-$\wr$-C-- group present in n-hexane, 4.3~eV for a C--H bond, and 4.4~eV for an allyl --C-$\wr$-C=C-$\wr$-C-- group \citep{Jones_Ysard_2025}. For comparison the National Institute of Standards and Technology (NIST) Computational Chemistry Comparison
and Benchmark DataBase (CCCBDB) bond energies of aromatic C=C in benzene are 5.4~eV, olefinic C=C in cyclohexene are 6.3~eV. The common alkynic C$\equiv$C bond energy is around 8.7~eV. Based on these values, the --C-$\wr$-C--C=C-- bonds are the most fragile and will be the first to break. Most of these allyl-adjacent groups can be found in the hydrogenated aromatic rings present in the domains. The UV photon-driven fragmentation will generate aromatic units linked by aliphatics. If these domains have a curvature they can form fullerenes by photoinduced dehydrogenation \citep{Irle_2006,Micelotta_2012}.
Alternatively, the UV field will break the aliphatic links and release the aromatic domains. If the domains have significant curvature, due to the presence of five--membered rings, they can close and form fullerenes \citep{Berne_2012, Berne_2015}. UV processing can further convert the defects (five/seven rings) into six-membered rings and form PAHs. 

Another fragmentation mechanism could involve multi-cation induced Coulomb fragmentation (MCF) due to the repulsion of charges localized on the aromatic domains. \citet{Jones_Ysard_2025} estimated that molecular particles with radii of 0.2 and 0.25~nm would contain on average two aromatic domains and that each domain would have to carry a charge of +2/+3 to induce fragmentation. 
Our molecular particle has a radius of 0.55~nm and contains six aromatic domains. This could lead to a total charge of +12 or more that is higher than the charges found on interstellar grains \citep{Ibanez-Mejia_2019} and this process is therefore unlikely. 

\section{Conclusions}
\label{sec:conclusions}
We have selected a potential hydrogenated carbonaceous molecular particle containing 130 carbon atoms, described its structure, computed its emission spectra for the 3.2--3.6~\micron,  5--13.7~\micron, and 15.0--22.0~\micron\ regions and compared it to the NIRSpec and MIRI/MRS spectra of IRAS 05110. The structure contains aromatic/hydrogenated aromatic domains with some defects (pairs of five/seven aromatic rings) and these domains are linked by aliphatic links. It has some curvature, due to the presence of aromatic five-membered rings, which reduces the particle's radius. The computed spectra are in qualitative agreement with the observed spectra. We have attributed the broad plateau, previously unassigned, to combinations of aliphatic, aromatic, and olefinic C--H bending modes that are coupled with the collective skeletal C--C modes of the disordered and distorted nature of these molecular particles. The UV photon-driven fragmentation of hydrogenated carbonaceous molecular particles can lead to the formation of PAHs and fullerenes and help explain the life-cycle of carbon in the Universe.

\section{acknowledgments}
This work is in part based on observations made with the NASA/ESA/CSA James Webb Space Telescope. All of the data presented in this article were obtained from the Mikulski Archive for Space Telescopes (MAST) at the Space Telescope Science Institute. These observations are associated with program \#4678 (DOI: 10.17909/zs5n-2t21).
E.P., J.C., N.C., and C.B. acknowledge support from the Canadian Space Agency (CSA, 24JWGO3B01), and the Natural Sciences and Engineering Research Council of Canada.
Support for US investigators (G.C.S., R.S.) in program \#4678 was provided by NASA through a grant from the Space Telescope Science Institute, which is operated by the Association of Universities for Research in Astronomy, Inc., under NASA contract NAS 5-03127.
A.R. acknowledges support from the Internal Scientist Funding Model (ISFM) Laboratory Astrophysics Directed Work Package Round 3 at NASA Ames and the High-End Computing (HEC) resources provided by the NASA Advanced Supercomputing (NAS) Division at the NASA Ames Research Center. G.C.S. was funded as part of this observing program by STScI through grant JWST-GO-04678.001-A. 
R.S.’s contribution to the research described here was carried out at the Jet Propulsion Laboratory, California Institute of Technology, under a contract with NASA (80NM0018D0004).
D.A.G.H. acknowledges support from the State Research Agency (AEI) of the Spanish Ministry of Science, Innovation, and Universities (MICIU) of the Government of Spain, and the European Regional Development fund (ERDF), under grant PID2023-147325NB-I00/AEI/10.13039/501100011033. This publication is based upon work from COST Action CA21126 - Carbon molecular nanostructures in space (NanoSpace), supported by COST (European Cooperation in Science and Technology).

\clearpage

\bibliographystyle{aasjournal}
\bibliography{aamnem99,bibliography}

\end{document}